\documentclass[%
 aip,
rsi,%
 amsmath,amssymb,
preprint,%
]{revtex4-1}
\usepackage{siunitx}
\usepackage{graphicx}
\usepackage{dcolumn}
\usepackage{bm}

\begin{document}

\preprint{}

\title[]{Design, fabrication and metrology of 10$\,\times\,$100 multi-planar integrated photonic routing manifolds for neural networks}

\author{Jeff Chiles}
 \email{jeffrey.chiles@nist.gov}
 \affiliation{ 
National Institute of Standards and Technology\\
Applied Physics Division\\ 
Boulder, CO, USA 80305
}%

\author{Sonia M. Buckley}%

 \affiliation{ 
National Institute of Standards and Technology\\
Applied Physics Division\\ 
Boulder, CO, USA 80305
}%

\author{Sae Woo Nam}

 \affiliation{ 
National Institute of Standards and Technology\\
Applied Physics Division\\ 
Boulder, CO, USA 80305
}%

\author{Richard P. Mirin}

 \affiliation{ 
National Institute of Standards and Technology\\
Applied Physics Division\\ 
Boulder, CO, USA 80305
}%

\author{Jeffrey M. Shainline}

 \affiliation{ 
National Institute of Standards and Technology\\
Applied Physics Division\\ 
Boulder, CO, USA 80305
}%

\date{\today}

\begin{abstract}
We design, fabricate and characterize integrated photonic routing manifolds with 10 inputs and 100 outputs using two vertically integrated planes of silicon nitride waveguides. We analyze manifolds via top-view camera imaging. This measurement technique allows the rapid acquisition of hundreds of precise transmission measurements.  We demonstrate manifolds with uniform and Gaussian power distribution patterns with mean power output errors (averaged over 10 sets of 10 inputs) of 0.7 and 0.9 dB, respectively, establishing this as a viable architecture for precision light distribution on-chip.  We also assess the performance of the passive photonic elements comprising the system via self-referenced test structures, including high-dynamic-range beam taps, waveguide cutback structures, and waveguide crossing arrays.
\end{abstract}

\maketitle

\section{\label{sec:Introduction}Introduction}
\subsection{\label{sec:Background}Background}
The development of highly compact and energy-efficient optical interconnects \cite{Miller2000} has been a major research objective for integrated photonics.  Applications of optical interconnects range from telecommunications \cite{zhang2017silicon} to energy-efficient and high-bandwidth cross-chip communications in CMOS systems \cite{suwa2015,sun2017}.  The reason for photonic communication to replace electrical communication is that light experiences no charge-based parasitics, and therefore can achieve higher fan-out as well as long-range communication with lower power and higher speed.  However, the relatively large size of photonic components presents challenges to their integration.  In a system with both photonic and electronic components, the chip area consumed by photonics grows rapidly as the number of communicating nodes, and their degree of connectivity, is increased.  For densely connected systems, the requisite number of waveguides can increase to the point where they cannot fit on one plane.  Wavelength-division-multiplexing (WDM)\cite{Zhou2009,sun2017} or mode-division multiplexing \cite{Jia2018} can partially alleviate this problem.  Only one or a small number of master communication buses is then required to satisfy the information bandwidth requirements.  When the number of nodes and their degree of connectivity is small, this provides an elegant and cost-effective solution to mitigating the von Neumann bottleneck \cite{Backus1978}.

However, neural computing departs significantly from the von Neumann architecture. In a neural system, each processing node (neuron) contains local memory and communicates to many other nodes of the network across local and global spatial scales \cite{sp2006,busp2009}. The information processing of biological neural systems is approximated in feed-forward neural networks, which have proven technologically useful \cite{Silver2016}.  A feed-forward neural network consists of multiple layers of neurons, which each integrate several inputs and transmit a signal when a threshold condition is reached.  Each layer consists of some number of neurons, which have directed connections to the next downstream layer of neurons. Computation and memory are distributed, largely eliminating the bottleneck of processor-memory communication, but necessitating significant communication to and from each neuron.

Light is naturally suited to perform this communication. Because photons are uncharged and massless, photons avoid charge-based wiring parasitics. Using light for communication in neural systems is very promising \cite{tana20142,shha2016,chri2017,tafe2017,Shainline2017,sh2018a}, but constructing a network of nodes each with thousands of connections presents a formidable routing challenge.  Using WDM alone is untenable, as it would require an extremely fine and precise wavelength spacing to be constantly maintained.  The ability to scale to greater connectivities thus depends on the number of waveguides that can be integrated on a substrate.  A suitable solution is the use of multiple planes of photonic waveguides, a field which has seen significant progress over the last decade \cite{Chiles2017,Sacher2017,Shang2015,Hosseinnia2015a,Kang2013,Seok2016}.  The stacking of waveguides allows for dense integration with low-loss and low-crosstalk waveguide crossings.  In the present work, we present the design and implementation of a two-plane signal distribution network routing 10 input nodes in one network layer to 100 connections on 10 output nodes. This routing manifold accomplishes the routing between two layers of a feed-forward neural network with 10 neurons per layer and all-to-all connectivity.  We recently reported a theoretical analysis of the performance and scaling of multi-planar routing strategies for neural computing \cite{sh2018b}.

\subsection{\label{sec:Design}Design}

\begin{figure}[t]
	\centering
	\fbox{\includegraphics[width=1.0\linewidth]{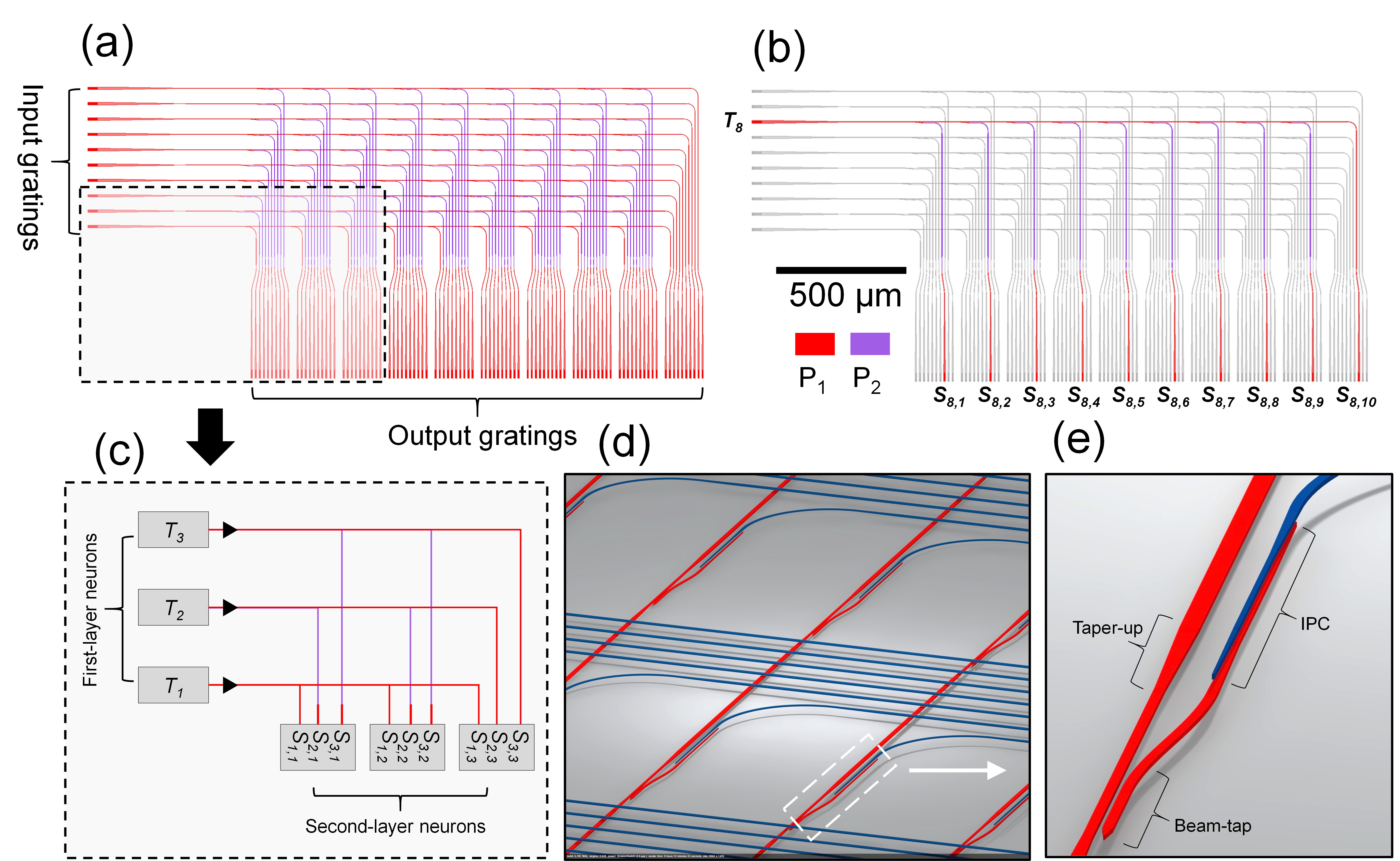}}
	\caption{Proposed photonic routing manifold design.  (a) Top-view of the schematic layout; (b) same view, with one transmission bus and associate output paths highlighted; (c) the representation of the boxed area in \textit{(a)} as a neural network, showing the notation scheme used throughout this work; (d) 3D perspective of the multi-planar system; (e) zoom view of the tap-and-transition device. IPC: inter-planar coupler; P\textsubscript{1} and P\textsubscript{2} refer to the bottom and top waveguide planes, respectively.}
	\label{fig:fig1b}
\end{figure}

Feed-forward neural networks commonly leverage topologies where a given layer has order $N^{2}$ synaptic connections, where $N$ is the number of neurons in a layer. In this work, we design, fabricate and experimentally characterize a distributed passive photonic routing manifold capable of realizing connectivities of order $N^{2}$. The routing network can be pruned to achieve any subset of connections. Communication with this manifold requires neither wavelength nor time multiplexing, yet can be straightforwardly extended to utilize either.   The design of the proposed manifold (Fig.\,\ref{fig:fig1b}) is based on two vertically integrated planes of waveguides.  The lower plane (P\textsubscript{1}) predominantly runs east, while the second plane (P\textsubscript{2}) runs south, thus avoiding in-plane crossings.  The light in P\textsubscript{1} bus waveguides originating from each input node is tapped sequentially into P\textsubscript{2} waveguides as the light propagates eastward.  This Manhattan-like routing architecture reduces the number of waveguides relative to a scheme where each input is immediately fanned-out with a star coupler.  

The manifold implements two layers of a feed-forward neural network with 10 upstream neurons (first layer), 10 downstream neurons with 100 synapses (second layer), and all-to-all connectivity. Figure\,\ref{fig:fig1b}(c) provides this perspective for a reduced section of the manifold.  Throughout the remainder of this paper, we will use the labeling scheme shown in Fig.\,\ref{fig:fig1b}(b-c): inputs (transmitters of the first layer of neurons) are denoted as $T_{x}$ and the receivers or outputs (synapses of the second layer neurons) are denoted as $S_{x,y}$.  For example, $S_{8,5}$ refers to the synapse on the fifth output neuron receiving input from the eighth input neuron, $T_{8}$.  The crossbar-like network allows each input node to be routed into a group of 10 outputs representing the whole input array (see the single-input case shown in Fig.\,\ref{fig:fig1b}(b)).  Each output group acts as the synapses (receivers) for that downstream neuron.  

The goal of the manifold is to route each input to one synapse on each output, following a pre-determined power distribution pattern. Here, we pursue two schemes to demonstrate control of the output intensity: uniform (each output synapse receives the same power) and Gaussian (the synapses from middle neurons of the upstream layer receive the most power, and the synapses from peripheral neurons receive much less).  A script was developed to automatically generate the layouts for the manifolds in both cases; variables in the script set neuron numbers as well as intensity distribution profiles.
The core element of the manifold is the tap-and-transition device shown in Fig.\,\ref{fig:fig1b}(e).  It comprises a beam-tap and an inter-planar coupler (IPC) in close proximity.  Its function is to divert a certain fraction of the bus power into a perpendicular waveguide on the upper plane.  Between bends, gratings, and tap-and-transition devices, the P\textsubscript{1} and P\textsubscript{2} waveguides are adiabatically tapered to and from a larger width (\SI{1.5}{\micro\meter}) to minimize scattering loss over most of their length.  The IPC is a similar design to the one presented in Ref. \onlinecite{Chiles2017}.  In the present work, the input waveguide (on P\textsubscript{1}) is tapered down to a width of 400 nm over a distance of \SI{12}{\micro\meter} and is then routed at a constant width for a distance of \SI{18}{\micro\meter}.  It is finally tapered down to a minimum width of 200 nm over \SI{12}{\micro\meter}.  The other waveguide (on P\textsubscript{2} receiving from P\textsubscript{1}) follows the same pattern in reverse over the same length.  The total IPC length is \SI{42}{\micro\meter}.

In a network of this size, significant dynamic range is required in the power-tap coefficients to achieve either uniform or Gaussian distributions.  If only a single coupling gap is utilized, two limits are encountered: (1) the finite size of the sine bend in the tap waveguide results in a certain minimum coupling coefficient, and (2) an excessively long interaction length is required to achieve a high coupling coefficient.  To address these issues, the manifold makes use of three coupling gaps and variable coupling lengths to improve the dynamic range of the power distribution network. The layout script selects the coupling gap from a look-up table generated from prior measurements of the tap coefficients. The three gap values are 300 nm, 400 nm, and 500 nm. Coupling lengths range from \SI{2.7}{\micro\meter} to \SI{19}{\micro\meter}.
\section{\label{sec:Experiment}Experiment}
\subsection{\label{sec:Fabrication}Fabrication}

\begin{figure}[t]
	\centering
	\fbox{\includegraphics[width=\linewidth]{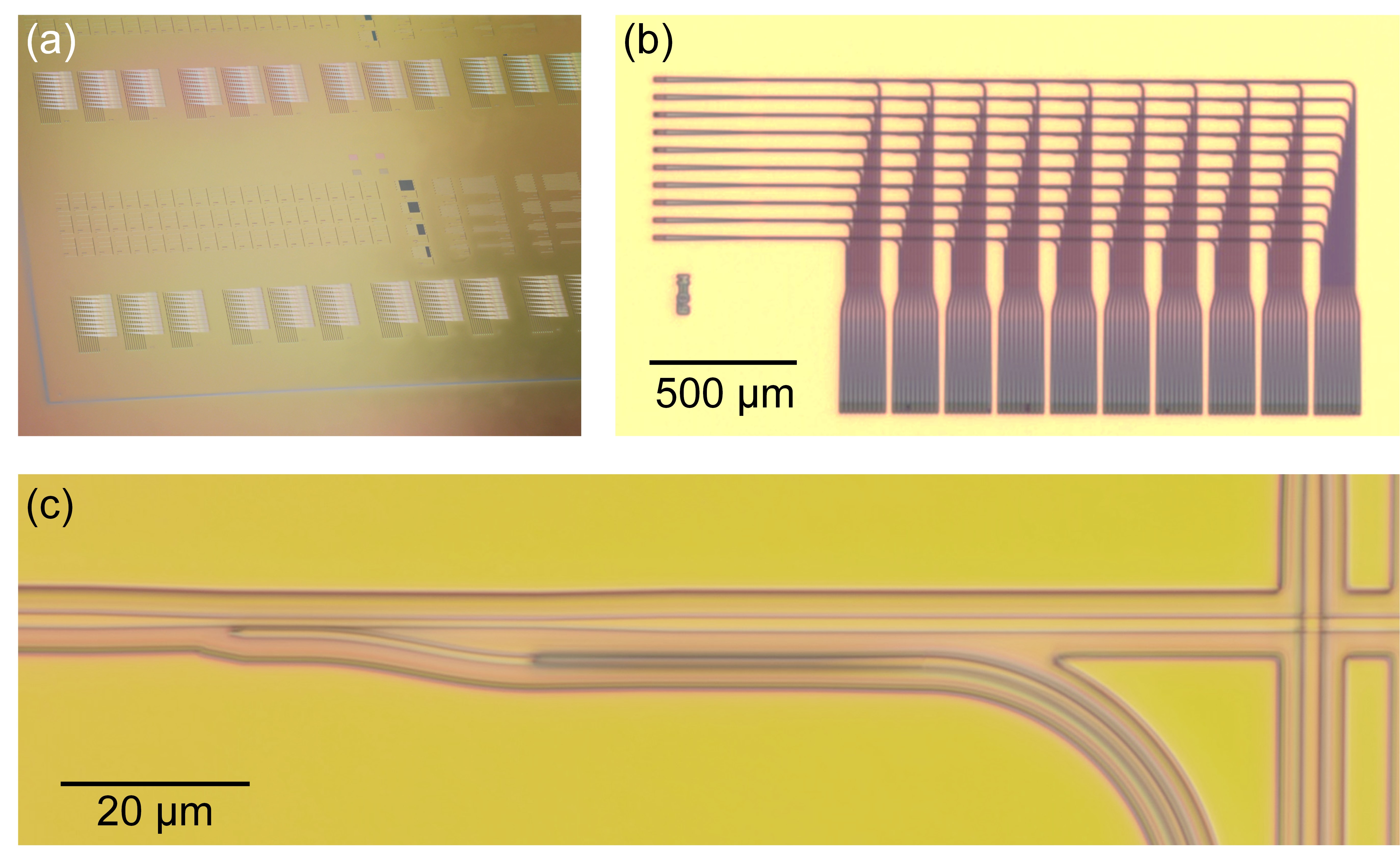}}
	\caption{Optical images of the fabricated devices: (a) focus-stacked view of several test devices on the wafer; (b) photonic routing manifold; and (c) zoom view of a beam-tap and interplanar coupler to move light from the P\textsubscript{1} to the P\textsubscript{2} plane.}
	\label{fig:fig2b}
\end{figure}

We fabricated the photonic routing manifolds at the NIST Boulder Microfabrication Facility on 76 mm silicon wafers.  Several images of the fabricated structures are shown in Fig.\,\ref{fig:fig2b}.  The fabrication process is similar to Ref.\,\onlinecite{Chiles2017}.  In the present work, the two waveguiding planes consist of 400 nm-thick silicon nitride (SiN), with an inter-planar pitch of approximately \SI{1.2}{\micro\meter} and a nominal width of 800 nm (on-mask).  The waveguides are cladded in plasma-deposited silicon dioxide (SiO$_{2}$) on all sides. The SiN material was deposited at a low temperature \cite{Shao2016,shainlineOE} of 40\,$^{\circ}$C to minimize stress from intrinsic sources and thermal expansion mismatch.  As such, the film was not optimized for propagation loss in this experiment, which was of little consequence over the short propagation lengths of the structures under test.  The SiN film exhibited a refractive index of 1.96 and a slab propagation loss of $\sim$5 dB per cm at $\lambda = 1310$ nm, measured via prism-coupling.

\subsection{\label{sec:Characterization}Characterization}
The manifolds under consideration each have 10 input ports and 100 output ports.  While it is possible to measure these devices with the common approach of aligning optical fibers to grating couplers or facet-terminated waveguides, that measurement technique has significant limitations.  First, repeatability strongly depends on the operator's ability to consistently optimize the fiber position on both ends using micro-positioning stages.  Second, sample and fiber position drift are likely to disturb any power normalization by the time all the output ports are measured.  V-groove arrays of fibers may alleviate the problem, but cannot accommodate densely packed structures, nor can the inter-fiber spacing be readily adjusted for different device configurations.  Realizing precise fiber array alignment to the sub-dB level is challenging.

\begin{figure}[t]
	\centering
	\fbox{\includegraphics[width=0.85\linewidth]{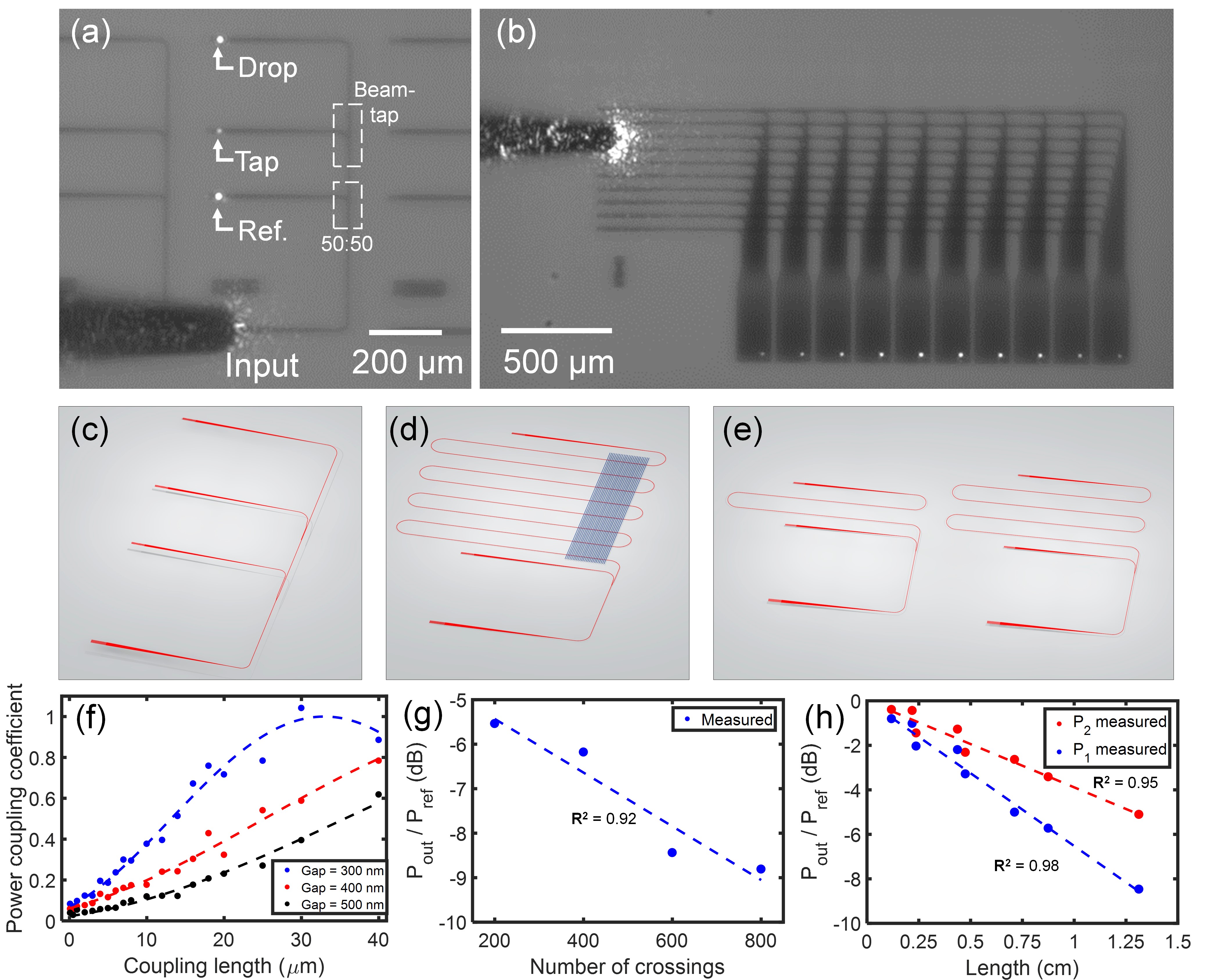}}
	\caption{(a) Infrared top-view image of a test structure for measuring a beam-tap.  The reference port allows immediate normalization of the data in the other two ports; (b) infrared top-view image of a Gaussian-distribution photonic routing manifold.  The bright spots are light emitted from the output grating couplers on the connected synapses; (c--e) 3D perspective views of the beam-tap, waveguide crossing and cut-back test structures, respectively; (f) measured beam-tap coefficients for three different coupling gaps used in the high-dynamic range power distribution system for the manifolds; (g) measured data for the P\textsubscript{1}/P\textsubscript{2} waveguide crossings, showing $6 \pm 1$ mdB loss per crossing; (h) measured waveguide cutback data, showing $6.5 \pm 0.4$ and $3.9 \pm 0.4$ dB per cm propagation loss for the P\textsubscript{1} and P\textsubscript{2} waveguides, respectively.  In (f--h), circles represent measured data points, and dashed lines indicate fits to the data.}
	\label{fig:fig3b}
\end{figure}

Here we pursue an alternative method of transmission measurements for this experiment: top-view imaging with a microscope and a camera. We couple transverse-electric (TE) polarized laser light near $\lambda = 1320$ nm onto the chip through a fiber-to-waveguide grating coupler, and light is coupled out through one or more grating couplers designed for vertical emission.  Instead of collecting the light with fibers, we focus it onto a 640\,$\times$\,512 pixel, 12-bit-depth indium gallium arsenide image sensor array through a microscope objective.  The light from each output port is integrated over a small window and normalized to the brightest port in the frame, allowing simultaneous acquisition of many outputs.  For most of the devices, a reference port is included near the input to allow straightforward normalization of the input power.  An \textit{in situ} image of this arrangement is shown in Fig.\,\ref{fig:fig3b}(a).  To obtain low-noise and repeatable measurements, we take care to meet several conditions during all measurements: (1) the camera's gamma (intensity curve) is always fixed at 1.0 to ensure linear power dependence and no gain is applied, (2) a pixel correction mask is applied to remove bright pixels and nonuniformities, (3) background light is filtered out via an 1150 nm long-pass filter inserted in the microscope tube, and (4) all output ports utilize an identical grating design and orientation. Proximity effect correction is applied during lithography to prevent distortion of the gratings in densely loaded areas. Any measurements with saturated pixels are rejected and repeated at a lower exposure time.  Likewise, measurements that are too close to the noise floor are repeated at a higher exposure time.  The grating coupler efficiency was not characterized, but it is more than sufficient to conduct the measurements with a high SNR.  Most measurements were conducted with a laser power of only a few hundred microwatts exiting the input fiber.

Images from the camera are analyzed with in-house software, which locates the optical modes of the output ports, and extracts a relative power measurement from the set.  The data analysis proceeds as follows: (1) Pixels corresponding to scattered light from the input fiber are set to zero; (2) the mean background intensity is subtracted from all pixels; (3) pixels with negative values are set to zero; (4) a convolution filter with a three-pixel by three-pixel window is applied to locate bright spots; and (5) power is integrated near each port, and the integration window is expanded until convergence to a specified residual is achieved.  The output of the script is an array of power values, normalized to the largest value in the set.

This measurement technique allows many photonic devices to be analyzed in parallel with high precision.  In this work, we investigate up to 10 ports at once, but many more ports can be analyzed, limited mainly by the imaging performance of the optics and camera which dictate some minimum spacing between ports.   Consider the test device in Fig.\,\ref{fig:fig3b}(a), which starts with an input grating coupler.  Light is then split into two paths in a 50:50 power splitter (based on a \textit{Y}-junction).  The path on the left leads to a reference output grating coupler.  On the right, the path leads to the device under test, in this case a beam-tap.  The coupling coefficient of the beam-tap is simply the ratio of the tap output power divided by the reference port's power.  The loss of the grating couplers, input waveguide section, and 50:50 splitter are normalized out. Consequently, the measurement has high throughput (fully parallel measurement of many ports) and is robust to alignment errors.  Most structures reported in this work, with the exception of the manifolds, were designed with this configuration.  In the case of the manifolds, the output ports (synapses) for a given input are measured relative to each other.
\subsubsection{\label{sec:Passive components}Passive components}
First, we characterized the performance of the different passive components that are used in the manifold.  The most critical feature is the high-dynamic-range power distribution system.  To analyze the constituent components, we measure an array of beam-tap test devices.  Across the array, the three coupling gaps of 300 nm, 400 nm, and 500 nm are implemented with a variety of coupling lengths.  Each test device comprises a 50:50 splitter and reference port followed by two device output ports: the tap output, and the drop output (indicating the untapped power).  The measured data are plotted in Fig.\,\ref{fig:fig3b}(f), along with a sine-squared fit of the coupling coefficient to the coupling length.  A tight fit is observed for all three coupling gaps, providing a reliable model for future routing manifold designs based on the same platform.  

Next, we analyzed the performance of the P\textsubscript{1}/P\textsubscript{2} waveguide crossings.  The distribution of these crossings is not uniform in the manifold design presented here, so some waveguides experience more crossing loss than others.  The $T_{1}$ bus waveguide (Fig.\,\ref{fig:fig1b}(a,c)) encounters 81 crossings, the maximum in this design.  A test structure for waveguide crossings is shown in Fig.\,\ref{fig:fig3b}(d).  It consists of a meandered P\textsubscript{1} waveguide passing under a cluster of P\textsubscript{2} waveguides above.  It crosses the P\textsubscript{2} waveguide cluster a total of 8 times.  Test structures with a total of 200, 400, 600 and 800 crossings were measured (Fig.\,\ref{fig:fig3b}(g)).  The P\textsubscript{2} waveguides are 800 nm wide (same width as the P\textsubscript{1} waveguides) and are spaced by a nominal period of \SI{4}{\micro\meter}, with a random variation between $\pm$ 400 nm to ensure no grating effects are introduced.  The data are fit with linear regression to a loss of 6 $\pm$ 1 mdB per crossing.  Considering the worst case of 81 crossings (path $S_{1,10}$), this constitutes a maximum link loss contribution of 0.49 dB.  In the manifolds presented later in this work, waveguide crossings occur between 1500 nm-wide waveguides, which may have slightly lower crossing losses due to tighter optical confinement; nevertheless, this measurement places a conservative bound on the loss value.

Waveguide propagation loss is also important to consider when trying to fabricate precision routing manifolds.  Cutback test structures are shown in Fig.\,\ref{fig:fig3b}(e).  Eight different path lengths between 1.2 to 13.0 mm were tested and identical structures were fabricated for both the P\textsubscript{1} and P\textsubscript{2} planes.  The data are shown in Fig.\,\ref{fig:fig3b}(h).  A good fit via linear regression is again observed, indicating propagation losses of $6.5 \pm 0.4$ and $3.9 \pm 0.4$ dB per cm, for the P\textsubscript{1} and P\textsubscript{2} waveguides, respectively.  The higher P\textsubscript{1} loss could be from mechanical degradation of its top oxide cladding in successive processing steps, which can be addressed with dense and robust sputtered oxide films. Future studies will include co-optimization of the optical and material properties of the SiN film to enable scaling to larger numbers of waveguiding planes.

Finally, we discuss the characterization of the IPCs.  On this mask, the IPC test structures were placed too far from the optimal zone in the middle of the wafer (where the planarization was on-target) resulting in a larger inter-planar pitch and higher than anticipated losses.  Since there were 64 IPCs back-to-back, the total loss exceeded the dynamic range possible in the measurement.  Fortunately, the IPC performance could still be straightforwardly characterized by comparing power transmission through two particular synapses on the manifolds: $S_{1,2}$ and $S_{2,2}$.  The only difference between them is that $S_{2,2}$ has two IPCs and \SI{180}{\micro\meter} extra P\textsubscript{1} propagation length.  We carefully aligned the fiber to each of the two inputs and recorded the power transmitted through the respective synapse.  At $\lambda=1320$ nm (the nominal wavelength for most tests in this work), a value of 0.6 dB per IPC is measured (after subtracting the 0.1 dB loss acquired from the extra propagation length).  This is sufficiently low loss to enable good power uniformity, since any two synapses may differ only by up to two IPCs in their routed paths.  Still, the loss is higher than anticipated, probably due to a deviation in the fabricated dimensions from the design.  In future work, we expect pre-compensation information to improve this to levels similar to our previous work on amorphous silicon \cite{Chiles2017}.    At this point, we can also make an informed estimate of the total link loss experienced in two representative paths through the manifold. First, we consider the path $S_{2,9}$ (Fig.\,\ref{fig:fig1b}), which encounters a relatively large loss compared to the other connections.  It has a long propagation length (2.9 mm, all on P\textsubscript{1}), 72 waveguide crossings, and 2 IPCs.  Utilizing the information collected from the passive measurements earlier in this section, we estimate the $S_{2,9}$ link loss to be 3.5 dB.  The smallest link loss occurs on $S_{1,1}$, which consists of 1.1 mm of P\textsubscript{1} propagation length, leading to 0.7 dB loss.  However, it should be noted that these losses are probably larger than the actual values, because we have used the propagation loss value from an 800 nm-wide waveguide.  In reality, the manifolds employ 1500 nm-wide waveguides over most of the propagation length, which could reduce the link loss in the $S_{2,9}$ case.
\begin{figure}[t]
	\centering
	\fbox{\includegraphics[width=0.8\linewidth]{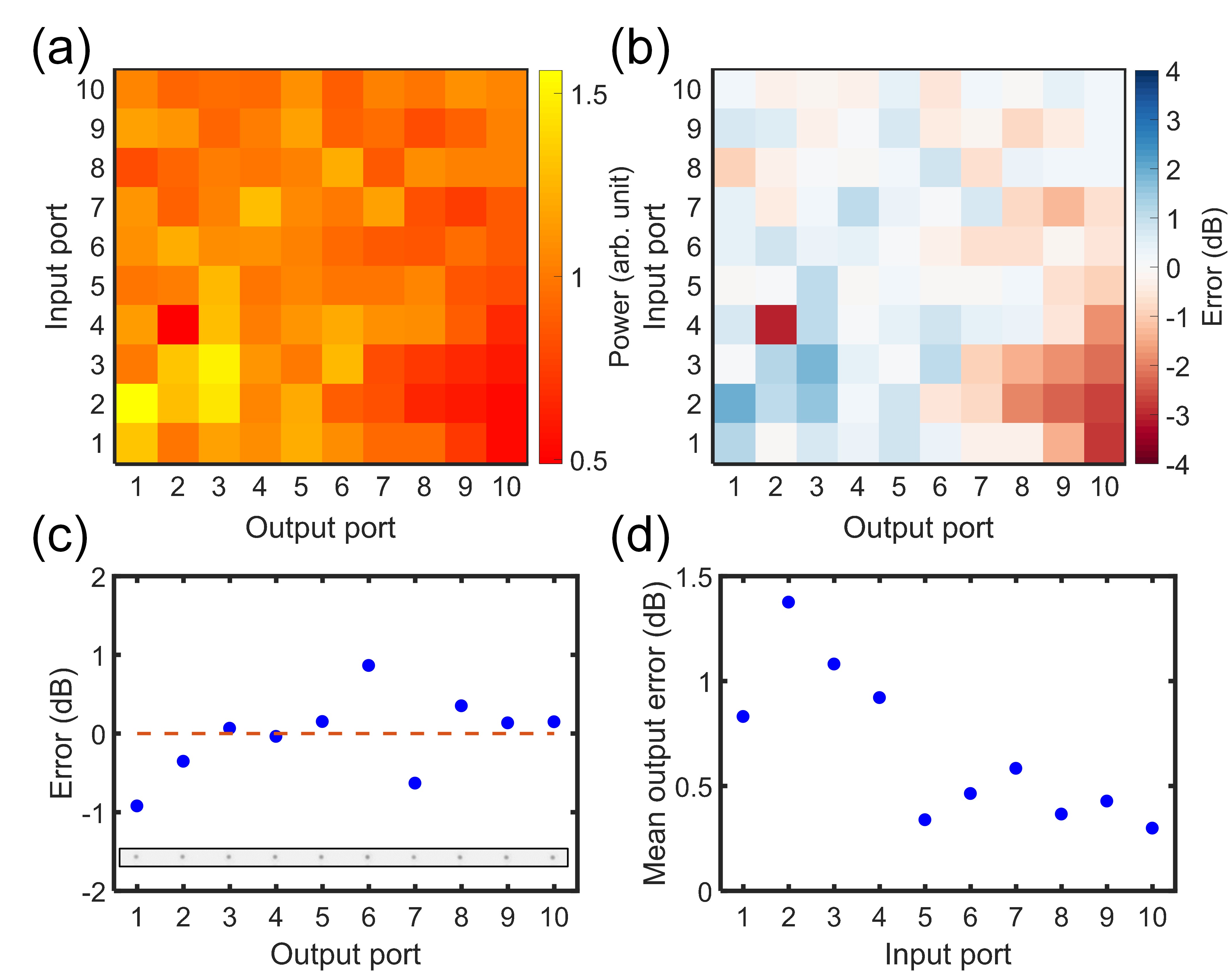}}
	\caption{Measured data for the uniform-distribution manifold measured at 1320 nm; (a) power distribution for all 100 synapses (each row is normalized to the data in its source microscope image); (b) error plot for the same data.  Error is calculated as the deviation of a synapse's value from the mean for that input; (c) error plotted for input $T_{8}$, showing the power on all synapses is within 1 dB of the mean, giving $<2$ dB spread.  The plot in (c) can be viewed as a horizontal slice of (a) for input $T_{8}$.  Inset: an infrared image of light from the ports on each synapse, with contrast enhanced for visibility; (d) the mean output error for each input case to observe trends in nonuniformity.  Higher error in the early inputs likely results from the higher density of crossings encountered by those waveguides.}
	\label{fig:fig4b}
\end{figure}

\subsubsection{\label{sec:Uniform-distribution manifold}Uniform-distribution manifold}

The first type of routing manifold we analyze is the uniform distribution pattern.  For any given input, the power delivered to each connected output synapse should be equal; for example, after applying input light to port $T_{x}$, we should observe a power distribution of $S_{x,1}=S_{x,2}=S_{x,3}\cdots=S_{x,10}$.  To satisfy this requirement, the tap coefficients range from 0.1 to 0.5.  An infrared image of the manifold under test is shown in Fig.\,\ref{fig:fig3b}(b), showing light emerging from the output ports.  The measured intensities (normalized for each input case) are plotted together in Fig.\,\ref{fig:fig4b}(a), as well as the errors in Fig.\,\ref{fig:fig4b}(b).  While there are a few outliers, the vast majority of synapses exhibit good uniformity.  The measured power uniformity of the outputs for input $T_{8}$ is shown in Fig.\,\ref{fig:fig4b}(c) as a representative case.  Error is calculated as the deviation of each point from the mean of that set.  In Fig.\,\ref{fig:fig4b}(d), the mean is calculated for the absolute value of the errors in each row in Fig.\,\ref{fig:fig4b}(b).  The grand mean of this data results in an overall average error of 0.7 dB.

\begin{figure}[t]
	\centering
	\fbox{\includegraphics[width=\linewidth]{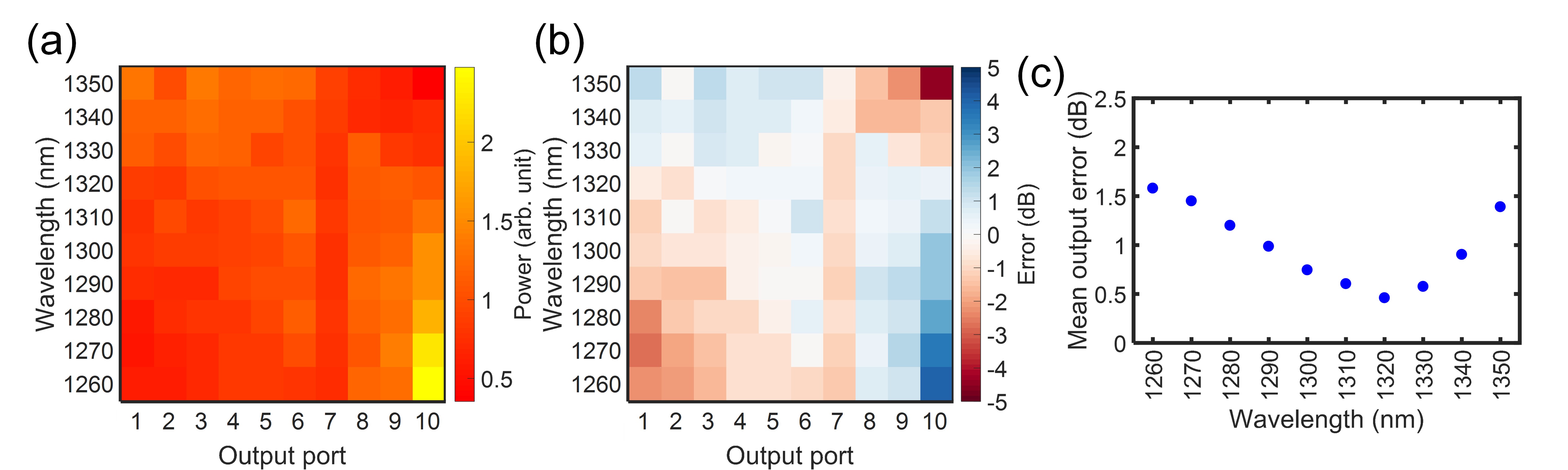}}
	\caption{Spectral performance for the uniform-distribution manifold; (a) output power for one input case ($T_{8}$) as the wavelength is varied; (b) resulting error plot calculated as the deviation from the mean; (c) mean error for each wavelength, showing optimal performance at $\lambda=1320$ nm.}
	\label{fig:fig5b}
\end{figure}

Next, we consider the spectral dependence of the uniform routing manifold.  For this study, we couple into a single input node $T_{8}$, and observe the changes to output uniformity while scanning the wavelength.  The power dependence on wavelength is plotted in Fig.\,\ref{fig:fig5b}(a), and the error in Fig.\,\ref{fig:fig5b}(b).  The lowest mean error of 0.46 dB is observed at a wavelength of 1320 nm (Fig.\,\ref{fig:fig5b}(c)), and the value remains below 1 dB over a bandwidth of at least 50 nm, providing sufficient tolerance for many applications.  We note that the mean error value only differs by 0.1 dB with the measurement of that same node, $T_{8}$, in the earlier series of measurements (see Fig.\,\ref{fig:fig4b}(d), input number 8). This indicates that the measurement approach is highly repeatable.
\subsubsection{\label{sec:Gaussian-distribution manifold}Gaussian-distribution manifold}
\begin{figure}[t]
	\centering
	\fbox{\includegraphics[width=0.8\linewidth]{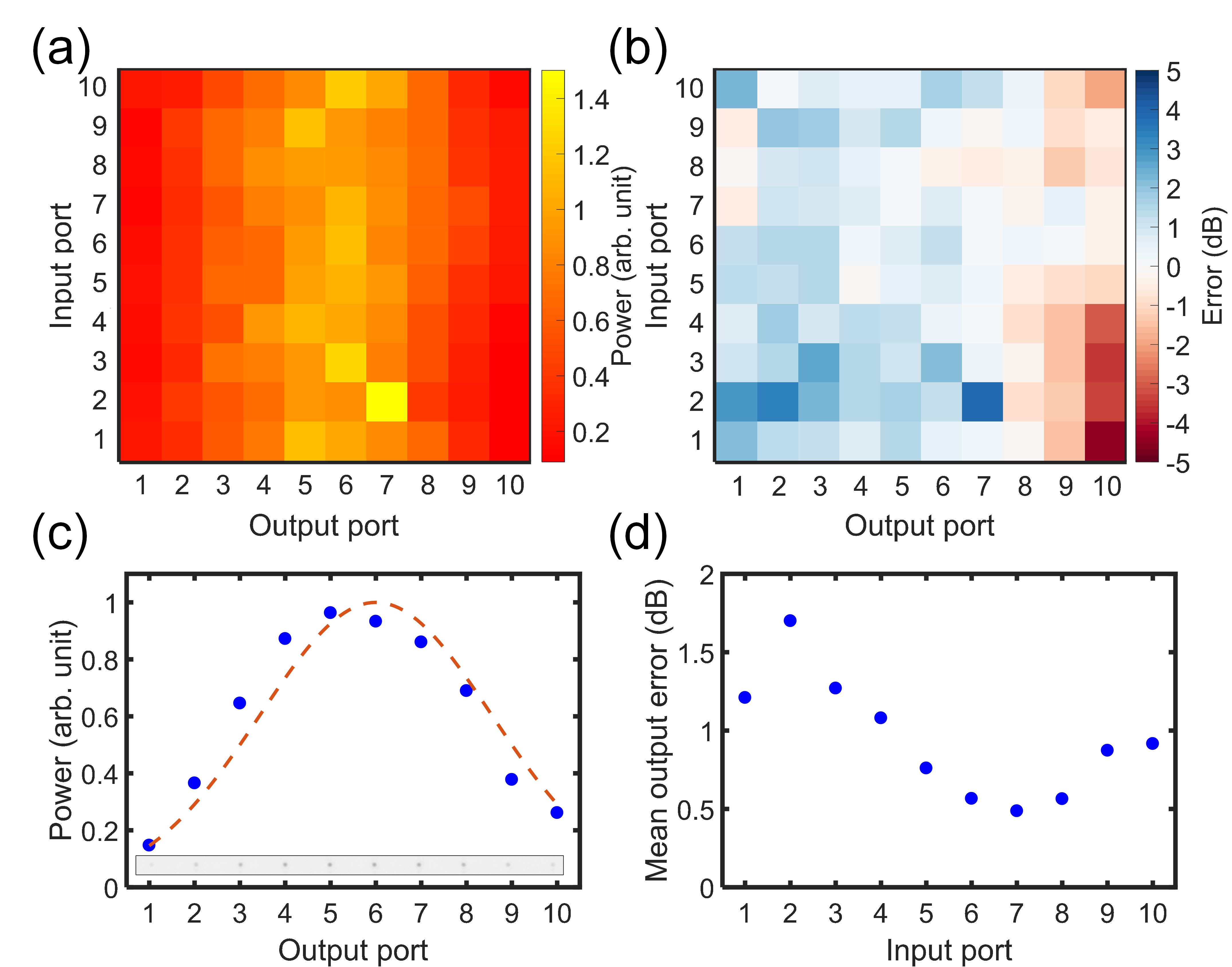}}
	\caption{Measured data for the Gaussian-distribution manifold measured at 1320 nm; (a) power distribution for all 100 synapses (each row is normalized to the data in that image); (b) error plot for the same data.  Error is calculated as the deviation of a synapse's value from the ideal Gaussian envelope that it was designed to follow (after fitting of the Gaussian amplitude and normalizing); (c) output powers plotted for the case of input $T_{8}$.  Circles are measured data, and the dashed line represents the designed Gaussian.  Inset: infrared image of light from each synapse, with contrast enhanced for visibility; (d) mean error for each input case.}
	\label{fig:fig6b}
\end{figure}

We continue the analysis with the Gaussian-distribution routing manifold.  This manifold is designed such that the synapses receive power following a Gaussian envelope.  The designed envelope is plotted on top of the experimentally measured synaptic power distribution for input node 8 in Fig.\,\ref{fig:fig6b}(c), showing good agreement.  The rest of the analysis follows the same pattern as for the uniform case.  Measured intensities are plotted together in Fig.\,\ref{fig:fig6b}(a), as well as the errors in Fig.\,\ref{fig:fig6b}(b).  For this manifold, the normalization for each input is done by least-squares fitting of the amplitude $a$ of the Gaussian power envelope $P(k)$ according to 
\begin{equation}
P(k)=ae^{\frac{-(4ln(2)(k-b)^{2})}{w^{2}}},
\end{equation}
where $k$ is the index of the output synapse, $b$ is the index of the peak value, and $w$ is the FWHM of the Gaussian envelope (both $b$ and $w$ are equal to 6 and are not fitted in the analysis).   Once $a$ is fitted, the output powers are normalized to that amplitude, so the envelopes remain in-line despite the occasional bright or dark synapse.  In Fig.\,\ref{fig:fig6b}(d), the mean is calculated for the absolute value of the errors in each row in Fig.\,\ref{fig:fig6b}(b).  The grand mean of the errors results in an overall average error of 0.9 dB.

\begin{figure}[t]
	\centering
	\fbox{\includegraphics[width=\linewidth]{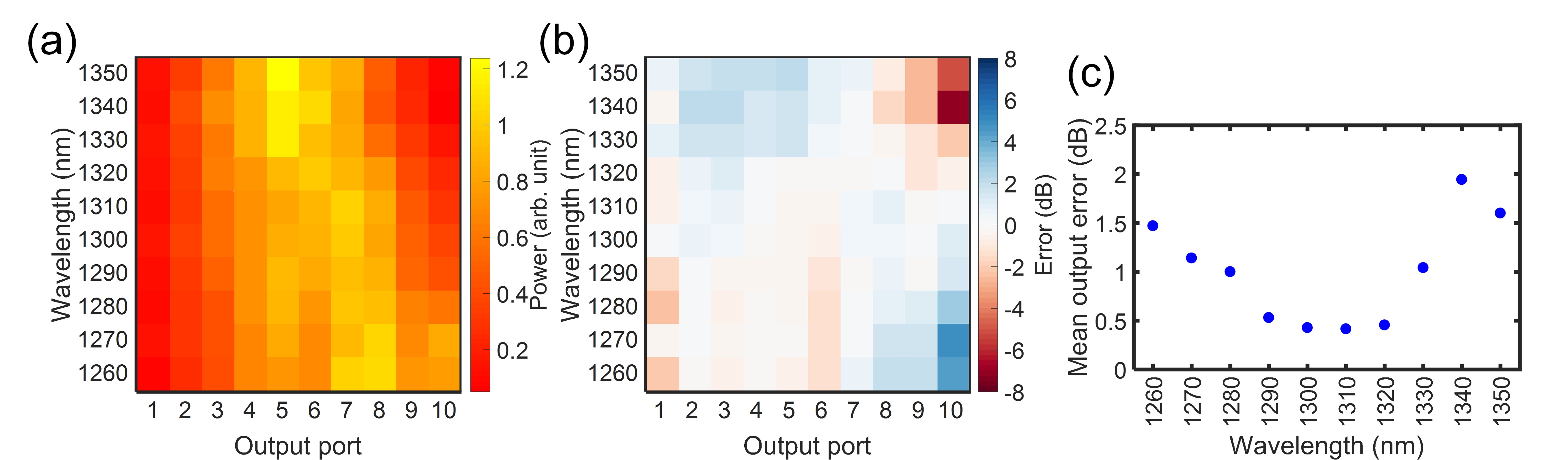}}
	\caption{Spectral performance for the Gaussian-distribution manifold; (a) output power for one input case ($T_{8}$) as the wavelength is varied; (b) resulting error plot calculated as the deviation from the ideal Gaussian envelope; (c) the mean error for each wavelength.}
	\label{fig:fig7b}
\end{figure}

The spectral dependence of the Gaussian routing manifold is analyzed with a similar method to the uniform manifold.  As before, light is coupled solely into $T_{8}$. The power dependence on wavelength is plotted in Fig.\,\ref{fig:fig7b}(a) and the error in Fig.\,\ref{fig:fig7b}(b).  A trend in the movement of the envelope's centroid toward lower-numbered synapses is seen in Fig.\,\ref{fig:fig7b}(a) as the wavelength is increased. This is consistent with the expectation that the coupling coefficients of the beam-taps will generally increase in the same direction.  The lowest error of 0.42 dB is observed at a wavelength of 1310 nm (Fig.\,\ref{fig:fig7b}(c)).  Since the uniform and Gaussian manifolds have the lowest error near a similar wavelength, we can conclude that the beam-tap coefficients are well-calibrated at 1310-1320 nm.

\section{\label{sec:Summary}Summary}

In this work, we propose, fabricate and characterize an integrated photonic routing manifold capable of distributing light with high precision across a 10$\,\times\,$100 network.  The approach utilizes multiple planes of waveguides and a distributed routing scheme to make efficient use of area.  The manifold can instantiate custom power distribution patterns, such as uniform or Gaussian, based on the values of beam tap coefficients. This design is topologically equivalent to a feed-forward, 10$\,\times\,$10, all-to-all-connected neural network. In analyzing the network and its sub-components, we employ a method for rapidly acquiring insertion loss measurements.  Using fiber-based input coupling and vertical grating emission onto an InGaAs imaging sensor, photonic routing manifolds with 100 output ports are fully characterized in less than 4 minutes.  At a wavelength of 1320 nm, the uniform and Gaussian manifolds were found to have mean output power errors (averaged over 10 rows of 10 inputs) of 0.7 and 0.9 dB, respectively.  These routing and measurement techniques offer new opportunities for complex integrated photonic systems in computing, telecommunications, and other applications.

Official contribution of the National Institute of Standards and Technology; not subject to copyright in the United States.

\nocite{*}
\bibliography{aipsamp}

\providecommand{\noopsort}[1]{}\providecommand{\singleletter}[1]{#1}%
\begin{thebibliography}{28}%
\makeatletter
\providecommand \@ifxundefined [1]{%
 \@ifx{#1\undefined}
}%
\providecommand \@ifnum [1]{%
 \ifnum #1\expandafter \@firstoftwo
 \else \expandafter \@secondoftwo
 \fi
}%
\providecommand \@ifx [1]{%
 \ifx #1\expandafter \@firstoftwo
 \else \expandafter \@secondoftwo
 \fi
}%
\providecommand \natexlab [1]{#1}%
\providecommand \enquote  [1]{``#1''}%
\providecommand \bibnamefont  [1]{#1}%
\providecommand \bibfnamefont [1]{#1}%
\providecommand \citenamefont [1]{#1}%
\providecommand \href@noop [0]{\@secondoftwo}%
\providecommand \href [0]{\begingroup \@sanitize@url \@href}%
\providecommand \@href[1]{\@@startlink{#1}\@@href}%
\providecommand \@@href[1]{\endgroup#1\@@endlink}%
\providecommand \@sanitize@url [0]{\catcode `\\12\catcode `\$12\catcode
  `\&12\catcode `\#12\catcode `\^12\catcode `\_12\catcode `\%12\relax}%
\providecommand \@@startlink[1]{}%
\providecommand \@@endlink[0]{}%
\providecommand \url  [0]{\begingroup\@sanitize@url \@url }%
\providecommand \@url [1]{\endgroup\@href {#1}{\urlprefix }}%
\providecommand \urlprefix  [0]{URL }%
\providecommand \Eprint [0]{\href }%
\providecommand \doibase [0]{http://dx.doi.org/}%
\providecommand \selectlanguage [0]{\@gobble}%
\providecommand \bibinfo  [0]{\@secondoftwo}%
\providecommand \bibfield  [0]{\@secondoftwo}%
\providecommand \translation [1]{[#1]}%
\providecommand \BibitemOpen [0]{}%
\providecommand \bibitemStop [0]{}%
\providecommand \bibitemNoStop [0]{.\EOS\space}%
\providecommand \EOS [0]{\spacefactor3000\relax}%
\providecommand \BibitemShut  [1]{\csname bibitem#1\endcsname}%
\let\auto@bib@innerbib\@empty
\bibitem [{\citenamefont {Miller}(2000)}]{Miller2000}%
  \BibitemOpen
  \bibfield  {author} {\bibinfo {author} {\bibfnamefont {D.}~\bibnamefont
  {Miller}},\ }\href {\doibase 10.1109/5.867687} {\bibfield  {journal}
  {\bibinfo  {journal} {Proceedings of the IEEE}\ }\textbf {\bibinfo {volume}
  {88}},\ \bibinfo {pages} {728} (\bibinfo {year} {2000})}\BibitemShut
  {NoStop}%
\bibitem [{\citenamefont {Zhang}\ and\ \citenamefont
  {Bowers}(2017)}]{zhang2017silicon}%
  \BibitemOpen
  \bibfield  {author} {\bibinfo {author} {\bibfnamefont {C.}~\bibnamefont
  {Zhang}}\ and\ \bibinfo {author} {\bibfnamefont {J.~E.}\ \bibnamefont
  {Bowers}},\ }\href@noop {} {\bibfield  {journal} {\bibinfo  {journal}
  {Optical Fiber Technology}\ } (\bibinfo {year} {2017})}\BibitemShut {NoStop}%
\bibitem [{\citenamefont {Sun}\ \emph {et~al.}(2015)\citenamefont {Sun},
  \citenamefont {Wade}, \citenamefont {Lee}, \citenamefont {Orcutt},
  \citenamefont {Alloatti}, \citenamefont {Georgas}, \citenamefont {Waterman},
  \citenamefont {Shainline}, \citenamefont {Avizienis}, \citenamefont {Lin},
  \citenamefont {Moss}, \citenamefont {Kumar}, \citenamefont {Pavanello},
  \citenamefont {Atabaki}, \citenamefont {Cook}, \citenamefont {Ou},
  \citenamefont {Leu}, \citenamefont {Chen}, \citenamefont {Asanovi{\' c}},
  \citenamefont {Ram}, \citenamefont {Popovi{\' c}},\ and\ \citenamefont
  {Stojanovi{\' c}}}]{suwa2015}%
  \BibitemOpen
  \bibfield  {author} {\bibinfo {author} {\bibfnamefont {C.}~\bibnamefont
  {Sun}}, \bibinfo {author} {\bibfnamefont {M.}~\bibnamefont {Wade}}, \bibinfo
  {author} {\bibfnamefont {Y.}~\bibnamefont {Lee}}, \bibinfo {author}
  {\bibfnamefont {J.}~\bibnamefont {Orcutt}}, \bibinfo {author} {\bibfnamefont
  {L.}~\bibnamefont {Alloatti}}, \bibinfo {author} {\bibfnamefont
  {M.}~\bibnamefont {Georgas}}, \bibinfo {author} {\bibfnamefont
  {A.}~\bibnamefont {Waterman}}, \bibinfo {author} {\bibfnamefont
  {J.}~\bibnamefont {Shainline}}, \bibinfo {author} {\bibfnamefont
  {R.}~\bibnamefont {Avizienis}}, \bibinfo {author} {\bibfnamefont
  {S.}~\bibnamefont {Lin}}, \bibinfo {author} {\bibfnamefont {B.}~\bibnamefont
  {Moss}}, \bibinfo {author} {\bibfnamefont {R.}~\bibnamefont {Kumar}},
  \bibinfo {author} {\bibfnamefont {F.}~\bibnamefont {Pavanello}}, \bibinfo
  {author} {\bibfnamefont {A.}~\bibnamefont {Atabaki}}, \bibinfo {author}
  {\bibfnamefont {H.}~\bibnamefont {Cook}}, \bibinfo {author} {\bibfnamefont
  {A.}~\bibnamefont {Ou}}, \bibinfo {author} {\bibfnamefont {J.}~\bibnamefont
  {Leu}}, \bibinfo {author} {\bibfnamefont {Y.-H.}\ \bibnamefont {Chen}},
  \bibinfo {author} {\bibfnamefont {K.}~\bibnamefont {Asanovi{\' c}}}, \bibinfo
  {author} {\bibfnamefont {R.}~\bibnamefont {Ram}}, \bibinfo {author}
  {\bibfnamefont {M.}~\bibnamefont {Popovi{\' c}}}, \ and\ \bibinfo {author}
  {\bibfnamefont {V.}~\bibnamefont {Stojanovi{\' c}}},\ }\href@noop {}
  {\bibfield  {journal} {\bibinfo  {journal} {Nature}\ }\textbf {\bibinfo
  {volume} {528}},\ \bibinfo {pages} {534} (\bibinfo {year}
  {2015})}\BibitemShut {NoStop}%
\bibitem [{\citenamefont {Sun}\ \emph {et~al.}(2017)\citenamefont {Sun},
  \citenamefont {Wade}, \citenamefont {Lee}, \citenamefont {Orcutt},
  \citenamefont {Alloatti}, \citenamefont {Georgas}, \citenamefont {Waterman},
  \citenamefont {Shainline}, \citenamefont {Avizienis}, \citenamefont {Lin},
  \citenamefont {Moss}, \citenamefont {Kumar}, \citenamefont {Pavanello},
  \citenamefont {Atabaki}, \citenamefont {Cook}, \citenamefont {Ou},
  \citenamefont {Leu}, \citenamefont {Chen}, \citenamefont {Asanovic},
  \citenamefont {Ram}, \citenamefont {Popovi{\'{c}}},\ and\ \citenamefont
  {Stojanovi{\'{c}}}}]{sun2017}%
  \BibitemOpen
  \bibfield  {author} {\bibinfo {author} {\bibfnamefont {C.}~\bibnamefont
  {Sun}}, \bibinfo {author} {\bibfnamefont {M.~T.}\ \bibnamefont {Wade}},
  \bibinfo {author} {\bibfnamefont {Y.}~\bibnamefont {Lee}}, \bibinfo {author}
  {\bibfnamefont {J.~S.}\ \bibnamefont {Orcutt}}, \bibinfo {author}
  {\bibfnamefont {L.}~\bibnamefont {Alloatti}}, \bibinfo {author}
  {\bibfnamefont {M.~S.}\ \bibnamefont {Georgas}}, \bibinfo {author}
  {\bibfnamefont {A.~S.}\ \bibnamefont {Waterman}}, \bibinfo {author}
  {\bibfnamefont {J.~M.}\ \bibnamefont {Shainline}}, \bibinfo {author}
  {\bibfnamefont {R.~R.}\ \bibnamefont {Avizienis}}, \bibinfo {author}
  {\bibfnamefont {S.}~\bibnamefont {Lin}}, \bibinfo {author} {\bibfnamefont
  {B.~R.}\ \bibnamefont {Moss}}, \bibinfo {author} {\bibfnamefont
  {R.}~\bibnamefont {Kumar}}, \bibinfo {author} {\bibfnamefont
  {F.}~\bibnamefont {Pavanello}}, \bibinfo {author} {\bibfnamefont {A.~H.}\
  \bibnamefont {Atabaki}}, \bibinfo {author} {\bibfnamefont {H.~M.}\
  \bibnamefont {Cook}}, \bibinfo {author} {\bibfnamefont {A.~J.}\ \bibnamefont
  {Ou}}, \bibinfo {author} {\bibfnamefont {J.~C.}\ \bibnamefont {Leu}},
  \bibinfo {author} {\bibfnamefont {Y.~H.}\ \bibnamefont {Chen}}, \bibinfo
  {author} {\bibfnamefont {K.}~\bibnamefont {Asanovic}}, \bibinfo {author}
  {\bibfnamefont {R.~J.}\ \bibnamefont {Ram}}, \bibinfo {author} {\bibfnamefont
  {M.~A.}\ \bibnamefont {Popovi{\'{c}}}}, \ and\ \bibinfo {author}
  {\bibfnamefont {V.~M.}\ \bibnamefont {Stojanovi{\'{c}}}},\ }in\ \href@noop {}
  {\emph {\bibinfo {booktitle} {2017 Optical Fiber Communications Conference
  and Exhibition (OFC)}}}\ (\bibinfo {year} {2017})\ pp.\ \bibinfo {pages}
  {1--3}\BibitemShut {NoStop}%
\bibitem [{\citenamefont {Zhou}\ \emph {et~al.}(2009)\citenamefont {Zhou},
  \citenamefont {Djordjevic}, \citenamefont {Proietti}, \citenamefont {Ding},
  \citenamefont {Yoo}, \citenamefont {Amirtharajah},\ and\ \citenamefont
  {Akella}}]{Zhou2009}%
  \BibitemOpen
  \bibfield  {author} {\bibinfo {author} {\bibfnamefont {L.}~\bibnamefont
  {Zhou}}, \bibinfo {author} {\bibfnamefont {S.~S.}\ \bibnamefont
  {Djordjevic}}, \bibinfo {author} {\bibfnamefont {R.}~\bibnamefont
  {Proietti}}, \bibinfo {author} {\bibfnamefont {D.}~\bibnamefont {Ding}},
  \bibinfo {author} {\bibfnamefont {S.~J.~B.}\ \bibnamefont {Yoo}}, \bibinfo
  {author} {\bibfnamefont {R.}~\bibnamefont {Amirtharajah}}, \ and\ \bibinfo
  {author} {\bibfnamefont {V.}~\bibnamefont {Akella}},\ }\href {\doibase
  10.1007/s00339-009-5121-6} {\bibfield  {journal} {\bibinfo  {journal}
  {Applied Physics A}\ }\textbf {\bibinfo {volume} {95}},\ \bibinfo {pages}
  {1111} (\bibinfo {year} {2009})}\BibitemShut {NoStop}%
\bibitem [{\citenamefont {Jia}\ \emph {et~al.}(2018)\citenamefont {Jia},
  \citenamefont {Zhou}, \citenamefont {Fu}, \citenamefont {Ding}, \citenamefont
  {Zhang},\ and\ \citenamefont {Yang}}]{Jia2018}%
  \BibitemOpen
  \bibfield  {author} {\bibinfo {author} {\bibfnamefont {H.}~\bibnamefont
  {Jia}}, \bibinfo {author} {\bibfnamefont {T.}~\bibnamefont {Zhou}}, \bibinfo
  {author} {\bibfnamefont {X.}~\bibnamefont {Fu}}, \bibinfo {author}
  {\bibfnamefont {J.}~\bibnamefont {Ding}}, \bibinfo {author} {\bibfnamefont
  {L.}~\bibnamefont {Zhang}}, \ and\ \bibinfo {author} {\bibfnamefont
  {L.}~\bibnamefont {Yang}},\ }\href {\doibase 10.1364/OE.26.009740} {\bibfield
   {journal} {\bibinfo  {journal} {Optics Express}\ }\textbf {\bibinfo {volume}
  {26}},\ \bibinfo {pages} {9740} (\bibinfo {year} {2018})}\BibitemShut
  {NoStop}%
\bibitem [{\citenamefont {Backus}\ and\ \citenamefont
  {John}(1978)}]{Backus1978}%
  \BibitemOpen
  \bibfield  {author} {\bibinfo {author} {\bibfnamefont {J.}~\bibnamefont
  {Backus}}\ and\ \bibinfo {author} {\bibnamefont {John}},\ }\href {\doibase
  10.1145/359576.359579} {\bibfield  {journal} {\bibinfo  {journal}
  {Communications of the ACM}\ }\textbf {\bibinfo {volume} {21}},\ \bibinfo
  {pages} {613} (\bibinfo {year} {1978})}\BibitemShut {NoStop}%
\bibitem [{\citenamefont {Sporns}(2006)}]{sp2006}%
  \BibitemOpen
  \bibfield  {author} {\bibinfo {author} {\bibfnamefont {O.}~\bibnamefont
  {Sporns}},\ }\href@noop {} {\bibfield  {journal} {\bibinfo  {journal}
  {BioSystems}\ }\textbf {\bibinfo {volume} {85}},\ \bibinfo {pages} {55}
  (\bibinfo {year} {2006})}\BibitemShut {NoStop}%
\bibitem [{\citenamefont {Bullmore}\ and\ \citenamefont
  {Sporns}(2009)}]{busp2009}%
  \BibitemOpen
  \bibfield  {author} {\bibinfo {author} {\bibfnamefont {E.}~\bibnamefont
  {Bullmore}}\ and\ \bibinfo {author} {\bibfnamefont {O.}~\bibnamefont
  {Sporns}},\ }\href@noop {} {\bibfield  {journal} {\bibinfo  {journal} {Nature
  Reviews Neuroscience}\ }\textbf {\bibinfo {volume} {10}},\ \bibinfo {pages}
  {186} (\bibinfo {year} {2009})}\BibitemShut {NoStop}%
\bibitem [{\citenamefont {Silver}\ \emph {et~al.}(2016)\citenamefont {Silver},
  \citenamefont {Huang}, \citenamefont {Maddison}, \citenamefont {Guez},
  \citenamefont {Sifre}, \citenamefont {van~den Driessche}, \citenamefont
  {Schrittwieser}, \citenamefont {Antonoglou}, \citenamefont {Panneershelvam},
  \citenamefont {Lanctot}, \citenamefont {Dieleman}, \citenamefont {Grewe},
  \citenamefont {Nham}, \citenamefont {Kalchbrenner}, \citenamefont
  {Sutskever}, \citenamefont {Lillicrap}, \citenamefont {Leach}, \citenamefont
  {Kavukcuoglu}, \citenamefont {Graepel},\ and\ \citenamefont
  {Hassabis}}]{Silver2016}%
  \BibitemOpen
  \bibfield  {author} {\bibinfo {author} {\bibfnamefont {D.}~\bibnamefont
  {Silver}}, \bibinfo {author} {\bibfnamefont {A.}~\bibnamefont {Huang}},
  \bibinfo {author} {\bibfnamefont {C.~J.}\ \bibnamefont {Maddison}}, \bibinfo
  {author} {\bibfnamefont {A.}~\bibnamefont {Guez}}, \bibinfo {author}
  {\bibfnamefont {L.}~\bibnamefont {Sifre}}, \bibinfo {author} {\bibfnamefont
  {G.}~\bibnamefont {van~den Driessche}}, \bibinfo {author} {\bibfnamefont
  {J.}~\bibnamefont {Schrittwieser}}, \bibinfo {author} {\bibfnamefont
  {I.}~\bibnamefont {Antonoglou}}, \bibinfo {author} {\bibfnamefont
  {V.}~\bibnamefont {Panneershelvam}}, \bibinfo {author} {\bibfnamefont
  {M.}~\bibnamefont {Lanctot}}, \bibinfo {author} {\bibfnamefont
  {S.}~\bibnamefont {Dieleman}}, \bibinfo {author} {\bibfnamefont
  {D.}~\bibnamefont {Grewe}}, \bibinfo {author} {\bibfnamefont
  {J.}~\bibnamefont {Nham}}, \bibinfo {author} {\bibfnamefont {N.}~\bibnamefont
  {Kalchbrenner}}, \bibinfo {author} {\bibfnamefont {I.}~\bibnamefont
  {Sutskever}}, \bibinfo {author} {\bibfnamefont {T.}~\bibnamefont
  {Lillicrap}}, \bibinfo {author} {\bibfnamefont {M.}~\bibnamefont {Leach}},
  \bibinfo {author} {\bibfnamefont {K.}~\bibnamefont {Kavukcuoglu}}, \bibinfo
  {author} {\bibfnamefont {T.}~\bibnamefont {Graepel}}, \ and\ \bibinfo
  {author} {\bibfnamefont {D.}~\bibnamefont {Hassabis}},\ }\href {\doibase
  10.1038/nature16961} {\bibfield  {journal} {\bibinfo  {journal} {Nature}\
  }\textbf {\bibinfo {volume} {529}},\ \bibinfo {pages} {484} (\bibinfo {year}
  {2016})}\BibitemShut {NoStop}%
\bibitem [{\citenamefont {Tait}\ \emph {et~al.}(2014)\citenamefont {Tait},
  \citenamefont {Nahmias}, \citenamefont {Shastri},\ and\ \citenamefont
  {Prucnal}}]{tana20142}%
  \BibitemOpen
  \bibfield  {author} {\bibinfo {author} {\bibfnamefont {A.}~\bibnamefont
  {Tait}}, \bibinfo {author} {\bibfnamefont {M.}~\bibnamefont {Nahmias}},
  \bibinfo {author} {\bibfnamefont {B.}~\bibnamefont {Shastri}}, \ and\
  \bibinfo {author} {\bibfnamefont {P.}~\bibnamefont {Prucnal}},\ }\href@noop
  {} {\bibfield  {journal} {\bibinfo  {journal} {J. Lightwave Technol.}\
  }\textbf {\bibinfo {volume} {32}},\ \bibinfo {pages} {3427} (\bibinfo {year}
  {2014})}\BibitemShut {NoStop}%
\bibitem [{\citenamefont {Shen}\ \emph {et~al.}(2016)\citenamefont {Shen},
  \citenamefont {Harris}, \citenamefont {Skirlo}, \citenamefont {Prabhu},
  \citenamefont {Baehr-Jones}, \citenamefont {Hochberg}, \citenamefont {Sun},
  \citenamefont {Zhao}, \citenamefont {Larochelle}, \citenamefont {Englund},\
  and\ \citenamefont {Soljacic}}]{shha2016}%
  \BibitemOpen
  \bibfield  {author} {\bibinfo {author} {\bibfnamefont {Y.}~\bibnamefont
  {Shen}}, \bibinfo {author} {\bibfnamefont {N.}~\bibnamefont {Harris}},
  \bibinfo {author} {\bibfnamefont {S.}~\bibnamefont {Skirlo}}, \bibinfo
  {author} {\bibfnamefont {M.}~\bibnamefont {Prabhu}}, \bibinfo {author}
  {\bibfnamefont {T.}~\bibnamefont {Baehr-Jones}}, \bibinfo {author}
  {\bibfnamefont {M.}~\bibnamefont {Hochberg}}, \bibinfo {author}
  {\bibfnamefont {X.}~\bibnamefont {Sun}}, \bibinfo {author} {\bibfnamefont
  {S.}~\bibnamefont {Zhao}}, \bibinfo {author} {\bibfnamefont {H.}~\bibnamefont
  {Larochelle}}, \bibinfo {author} {\bibfnamefont {D.}~\bibnamefont {Englund}},
  \ and\ \bibinfo {author} {\bibfnamefont {M.}~\bibnamefont {Soljacic}},\
  }\href@noop {} {\bibfield  {journal} {\bibinfo  {journal} {Nature Photonics}\
  }\textbf {\bibinfo {volume} {11}},\ \bibinfo {pages} {441} (\bibinfo {year}
  {2016})}\BibitemShut {NoStop}%
\bibitem [{\citenamefont {Cheng}\ \emph {et~al.}(2017)\citenamefont {Cheng},
  \citenamefont {Rios}, \citenamefont {Pernice}, \citenamefont {Wright},\ and\
  \citenamefont {Bhaskaran}}]{chri2017}%
  \BibitemOpen
  \bibfield  {author} {\bibinfo {author} {\bibfnamefont {Z.}~\bibnamefont
  {Cheng}}, \bibinfo {author} {\bibfnamefont {C.}~\bibnamefont {Rios}},
  \bibinfo {author} {\bibfnamefont {W.}~\bibnamefont {Pernice}}, \bibinfo
  {author} {\bibfnamefont {C.}~\bibnamefont {Wright}}, \ and\ \bibinfo {author}
  {\bibfnamefont {H.}~\bibnamefont {Bhaskaran}},\ }\href@noop {} {\bibfield
  {journal} {\bibinfo  {journal} {Science Advances}\ }\textbf {\bibinfo
  {volume} {3}},\ \bibinfo {pages} {1700160} (\bibinfo {year}
  {2017})}\BibitemShut {NoStop}%
\bibitem [{\citenamefont {Tait}\ \emph {et~al.}(2017)\citenamefont {Tait},
  \citenamefont {de~Lima}, \citenamefont {Zhou}, \citenamefont {Wu},
  \citenamefont {Nahmias}, \citenamefont {Shastri},\ and\ \citenamefont
  {Prucnal}}]{tafe2017}%
  \BibitemOpen
  \bibfield  {author} {\bibinfo {author} {\bibfnamefont {A.}~\bibnamefont
  {Tait}}, \bibinfo {author} {\bibfnamefont {T.~F.}\ \bibnamefont {de~Lima}},
  \bibinfo {author} {\bibfnamefont {E.}~\bibnamefont {Zhou}}, \bibinfo {author}
  {\bibfnamefont {A.}~\bibnamefont {Wu}}, \bibinfo {author} {\bibfnamefont
  {M.}~\bibnamefont {Nahmias}}, \bibinfo {author} {\bibfnamefont
  {B.}~\bibnamefont {Shastri}}, \ and\ \bibinfo {author} {\bibfnamefont
  {P.}~\bibnamefont {Prucnal}},\ }\href@noop {} {\bibfield  {journal} {\bibinfo
   {journal} {Nature Sci. Rep.}\ }\textbf {\bibinfo {volume} {7}},\ \bibinfo
  {pages} {7430} (\bibinfo {year} {2017})}\BibitemShut {NoStop}%
\bibitem [{\citenamefont {Shainline}\ \emph
  {et~al.}(2017{\natexlab{a}})\citenamefont {Shainline}, \citenamefont
  {Buckley}, \citenamefont {Mirin},\ and\ \citenamefont {Nam}}]{Shainline2017}%
  \BibitemOpen
  \bibfield  {author} {\bibinfo {author} {\bibfnamefont {J.~M.}\ \bibnamefont
  {Shainline}}, \bibinfo {author} {\bibfnamefont {S.~M.}\ \bibnamefont
  {Buckley}}, \bibinfo {author} {\bibfnamefont {R.~P.}\ \bibnamefont {Mirin}},
  \ and\ \bibinfo {author} {\bibfnamefont {S.~W.}\ \bibnamefont {Nam}},\ }\href
  {\doibase 10.1103/PhysRevApplied.7.034013} {\bibfield  {journal} {\bibinfo
  {journal} {Physical Review Applied}\ }\textbf {\bibinfo {volume} {7}},\
  \bibinfo {pages} {034013} (\bibinfo {year} {2017}{\natexlab{a}})}\BibitemShut
  {NoStop}%
\bibitem [{\citenamefont {Shainline}\ \emph
  {et~al.}(2018{\natexlab{a}})\citenamefont {Shainline}, \citenamefont
  {Buckley}, \citenamefont {McCaughan}, \citenamefont {Chiles}, \citenamefont
  {Mirin},\ and\ \citenamefont {Nam}}]{sh2018a}%
  \BibitemOpen
  \bibfield  {author} {\bibinfo {author} {\bibfnamefont {J.~M.}\ \bibnamefont
  {Shainline}}, \bibinfo {author} {\bibfnamefont {S.~M.}\ \bibnamefont
  {Buckley}}, \bibinfo {author} {\bibfnamefont {A.~N.}\ \bibnamefont
  {McCaughan}}, \bibinfo {author} {\bibfnamefont {J.}~\bibnamefont {Chiles}},
  \bibinfo {author} {\bibfnamefont {R.~P.}\ \bibnamefont {Mirin}}, \ and\
  \bibinfo {author} {\bibfnamefont {S.~W.}\ \bibnamefont {Nam}},\ }\href
  {http://arxiv.org/abs/1805.01929} {\  (\bibinfo {year}
  {2018}{\natexlab{a}})},\ \Eprint {http://arxiv.org/abs/1805.01929}
  {arXiv:1805.01929} \BibitemShut {NoStop}%
\bibitem [{\citenamefont {Chiles}\ \emph {et~al.}(2017)\citenamefont {Chiles},
  \citenamefont {Buckley}, \citenamefont {Nader}, \citenamefont {Nam},
  \citenamefont {Mirin},\ and\ \citenamefont {Shainline}}]{Chiles2017}%
  \BibitemOpen
  \bibfield  {author} {\bibinfo {author} {\bibfnamefont {J.}~\bibnamefont
  {Chiles}}, \bibinfo {author} {\bibfnamefont {S.}~\bibnamefont {Buckley}},
  \bibinfo {author} {\bibfnamefont {N.}~\bibnamefont {Nader}}, \bibinfo
  {author} {\bibfnamefont {S.~W.}\ \bibnamefont {Nam}}, \bibinfo {author}
  {\bibfnamefont {R.~P.}\ \bibnamefont {Mirin}}, \ and\ \bibinfo {author}
  {\bibfnamefont {J.~M.}\ \bibnamefont {Shainline}},\ }\href {\doibase
  10.1063/1.5000384} {\bibfield  {journal} {\bibinfo  {journal} {APL
  Photonics}\ }\textbf {\bibinfo {volume} {2}},\ \bibinfo {pages} {116101}
  (\bibinfo {year} {2017})}\BibitemShut {NoStop}%
\bibitem [{\citenamefont {Sacher}\ \emph {et~al.}(2017)\citenamefont {Sacher},
  \citenamefont {Mikkelsen}, \citenamefont {Dumais}, \citenamefont {Jiang},
  \citenamefont {Goodwill}, \citenamefont {Luo}, \citenamefont {Huang},
  \citenamefont {Yang}, \citenamefont {Bois}, \citenamefont {Lo}, \citenamefont
  {Bernier},\ and\ \citenamefont {Poon}}]{Sacher2017}%
  \BibitemOpen
  \bibfield  {author} {\bibinfo {author} {\bibfnamefont {W.~D.}\ \bibnamefont
  {Sacher}}, \bibinfo {author} {\bibfnamefont {J.~C.}\ \bibnamefont
  {Mikkelsen}}, \bibinfo {author} {\bibfnamefont {P.}~\bibnamefont {Dumais}},
  \bibinfo {author} {\bibfnamefont {J.}~\bibnamefont {Jiang}}, \bibinfo
  {author} {\bibfnamefont {D.}~\bibnamefont {Goodwill}}, \bibinfo {author}
  {\bibfnamefont {X.}~\bibnamefont {Luo}}, \bibinfo {author} {\bibfnamefont
  {Y.}~\bibnamefont {Huang}}, \bibinfo {author} {\bibfnamefont
  {Y.}~\bibnamefont {Yang}}, \bibinfo {author} {\bibfnamefont {A.}~\bibnamefont
  {Bois}}, \bibinfo {author} {\bibfnamefont {P.~G.-Q.}\ \bibnamefont {Lo}},
  \bibinfo {author} {\bibfnamefont {E.}~\bibnamefont {Bernier}}, \ and\
  \bibinfo {author} {\bibfnamefont {J.~K.~S.}\ \bibnamefont {Poon}},\ }\href
  {\doibase 10.1364/OE.25.030862} {\bibfield  {journal} {\bibinfo  {journal}
  {Optics Express}\ }\textbf {\bibinfo {volume} {25}},\ \bibinfo {pages}
  {30862} (\bibinfo {year} {2017})}\BibitemShut {NoStop}%
\bibitem [{\citenamefont {Shang}\ \emph {et~al.}(2015)\citenamefont {Shang},
  \citenamefont {Pathak}, \citenamefont {Guan}, \citenamefont {Liu},\ and\
  \citenamefont {Yoo}}]{Shang2015}%
  \BibitemOpen
  \bibfield  {author} {\bibinfo {author} {\bibfnamefont {K.}~\bibnamefont
  {Shang}}, \bibinfo {author} {\bibfnamefont {S.}~\bibnamefont {Pathak}},
  \bibinfo {author} {\bibfnamefont {B.}~\bibnamefont {Guan}}, \bibinfo {author}
  {\bibfnamefont {G.}~\bibnamefont {Liu}}, \ and\ \bibinfo {author}
  {\bibfnamefont {S.~J.~B.}\ \bibnamefont {Yoo}},\ }\href {\doibase
  10.1364/OE.23.021334} {\bibfield  {journal} {\bibinfo  {journal} {Optics
  Express}\ }\textbf {\bibinfo {volume} {23}},\ \bibinfo {pages} {21334}
  (\bibinfo {year} {2015})}\BibitemShut {NoStop}%
\bibitem [{\citenamefont {Hosseinnia}\ \emph {et~al.}(2015)\citenamefont
  {Hosseinnia}, \citenamefont {Atabaki}, \citenamefont {Eftekhar},\ and\
  \citenamefont {Adibi}}]{Hosseinnia2015a}%
  \BibitemOpen
  \bibfield  {author} {\bibinfo {author} {\bibfnamefont {A.~H.}\ \bibnamefont
  {Hosseinnia}}, \bibinfo {author} {\bibfnamefont {A.~H.}\ \bibnamefont
  {Atabaki}}, \bibinfo {author} {\bibfnamefont {A.~A.}\ \bibnamefont
  {Eftekhar}}, \ and\ \bibinfo {author} {\bibfnamefont {A.}~\bibnamefont
  {Adibi}},\ }\href {\doibase 10.1364/OE.23.030297} {\bibfield  {journal}
  {\bibinfo  {journal} {Optics Express}\ }\textbf {\bibinfo {volume} {23}},\
  \bibinfo {pages} {30297} (\bibinfo {year} {2015})}\BibitemShut {NoStop}%
\bibitem [{\citenamefont {Kang}\ \emph {et~al.}(2013)\citenamefont {Kang},
  \citenamefont {Nishiyama}, \citenamefont {Atsumi}, \citenamefont {Amemiya},\
  and\ \citenamefont {Arai}}]{Kang2013}%
  \BibitemOpen
  \bibfield  {author} {\bibinfo {author} {\bibfnamefont {J.}~\bibnamefont
  {Kang}}, \bibinfo {author} {\bibfnamefont {N.}~\bibnamefont {Nishiyama}},
  \bibinfo {author} {\bibfnamefont {Y.}~\bibnamefont {Atsumi}}, \bibinfo
  {author} {\bibfnamefont {T.}~\bibnamefont {Amemiya}}, \ and\ \bibinfo
  {author} {\bibfnamefont {S.}~\bibnamefont {Arai}},\ }in\ \href
  {http://dx.doi.org/10.1117/12.2009456} {\emph {\bibinfo {booktitle}
  {Optoelectronic Interconnects XIII}}},\ Vol.\ \bibinfo {volume} {8630}\
  (\bibinfo {year} {2013})\ pp.\ \bibinfo {pages} {863008--863012}\BibitemShut
  {NoStop}%
\bibitem [{\citenamefont {Seok}\ \emph {et~al.}(2016)\citenamefont {Seok},
  \citenamefont {Quack}, \citenamefont {Han}, \citenamefont {Muller},\ and\
  \citenamefont {Wu}}]{Seok2016}%
  \BibitemOpen
  \bibfield  {author} {\bibinfo {author} {\bibfnamefont {T.~J.}\ \bibnamefont
  {Seok}}, \bibinfo {author} {\bibfnamefont {N.}~\bibnamefont {Quack}},
  \bibinfo {author} {\bibfnamefont {S.}~\bibnamefont {Han}}, \bibinfo {author}
  {\bibfnamefont {R.~S.}\ \bibnamefont {Muller}}, \ and\ \bibinfo {author}
  {\bibfnamefont {M.~C.}\ \bibnamefont {Wu}},\ }\href {\doibase
  10.1364/OPTICA.3.000064} {\bibfield  {journal} {\bibinfo  {journal} {Optica}\
  }\textbf {\bibinfo {volume} {3}},\ \bibinfo {pages} {64} (\bibinfo {year}
  {2016})}\BibitemShut {NoStop}%
\bibitem [{\citenamefont {Shainline}\ \emph
  {et~al.}(2018{\natexlab{b}})\citenamefont {Shainline}, \citenamefont
  {Chiles}, \citenamefont {Buckley}, \citenamefont {McCaughan}, \citenamefont
  {Mirin},\ and\ \citenamefont {Nam}}]{sh2018b}%
  \BibitemOpen
  \bibfield  {author} {\bibinfo {author} {\bibfnamefont {J.~M.}\ \bibnamefont
  {Shainline}}, \bibinfo {author} {\bibfnamefont {J.}~\bibnamefont {Chiles}},
  \bibinfo {author} {\bibfnamefont {S.~M.}\ \bibnamefont {Buckley}}, \bibinfo
  {author} {\bibfnamefont {A.~N.}\ \bibnamefont {McCaughan}}, \bibinfo {author}
  {\bibfnamefont {R.~P.}\ \bibnamefont {Mirin}}, \ and\ \bibinfo {author}
  {\bibfnamefont {S.~W.}\ \bibnamefont {Nam}},\ }\href
  {http://arxiv.org/abs/1805.01942} {\  (\bibinfo {year}
  {2018}{\natexlab{b}})},\ \Eprint {http://arxiv.org/abs/1805.01942}
  {arXiv:1805.01942} \BibitemShut {NoStop}%
\bibitem [{\citenamefont {Shao}\ \emph {et~al.}(2016)\citenamefont {Shao},
  \citenamefont {Chen}, \citenamefont {Chen}, \citenamefont {Zhang},
  \citenamefont {Zhang}, \citenamefont {Jian}, \citenamefont {Fan},
  \citenamefont {Liu}, \citenamefont {Yang}, \citenamefont {Zhou},\ and\
  \citenamefont {Yu}}]{Shao2016}%
  \BibitemOpen
  \bibfield  {author} {\bibinfo {author} {\bibfnamefont {Z.}~\bibnamefont
  {Shao}}, \bibinfo {author} {\bibfnamefont {Y.}~\bibnamefont {Chen}}, \bibinfo
  {author} {\bibfnamefont {H.}~\bibnamefont {Chen}}, \bibinfo {author}
  {\bibfnamefont {Y.}~\bibnamefont {Zhang}}, \bibinfo {author} {\bibfnamefont
  {F.}~\bibnamefont {Zhang}}, \bibinfo {author} {\bibfnamefont
  {J.}~\bibnamefont {Jian}}, \bibinfo {author} {\bibfnamefont {Z.}~\bibnamefont
  {Fan}}, \bibinfo {author} {\bibfnamefont {L.}~\bibnamefont {Liu}}, \bibinfo
  {author} {\bibfnamefont {C.}~\bibnamefont {Yang}}, \bibinfo {author}
  {\bibfnamefont {L.}~\bibnamefont {Zhou}}, \ and\ \bibinfo {author}
  {\bibfnamefont {S.}~\bibnamefont {Yu}},\ }\href {\doibase
  10.1364/OE.24.001865} {\bibfield  {journal} {\bibinfo  {journal} {Optics
  Express}\ }\textbf {\bibinfo {volume} {24}},\ \bibinfo {pages} {1865}
  (\bibinfo {year} {2016})}\BibitemShut {NoStop}%
\bibitem [{\citenamefont {Shainline}\ \emph
  {et~al.}(2017{\natexlab{b}})\citenamefont {Shainline}, \citenamefont
  {Buckley}, \citenamefont {Nader}, \citenamefont {Gentry}, \citenamefont
  {Cossel}, \citenamefont {Cleary}, \citenamefont {Popovi{\' c}}, \citenamefont
  {Newbury}, \citenamefont {Nam},\ and\ \citenamefont {Mirin}}]{shainlineOE}%
  \BibitemOpen
  \bibfield  {author} {\bibinfo {author} {\bibfnamefont {J.~M.}\ \bibnamefont
  {Shainline}}, \bibinfo {author} {\bibfnamefont {S.~M.}\ \bibnamefont
  {Buckley}}, \bibinfo {author} {\bibfnamefont {N.}~\bibnamefont {Nader}},
  \bibinfo {author} {\bibfnamefont {C.~M.}\ \bibnamefont {Gentry}}, \bibinfo
  {author} {\bibfnamefont {K.~C.}\ \bibnamefont {Cossel}}, \bibinfo {author}
  {\bibfnamefont {J.~W.}\ \bibnamefont {Cleary}}, \bibinfo {author}
  {\bibfnamefont {M.}~\bibnamefont {Popovi{\' c}}}, \bibinfo {author}
  {\bibfnamefont {N.~R.}\ \bibnamefont {Newbury}}, \bibinfo {author}
  {\bibfnamefont {S.~W.}\ \bibnamefont {Nam}}, \ and\ \bibinfo {author}
  {\bibfnamefont {R.~P.}\ \bibnamefont {Mirin}},\ }\href {\doibase
  10.1364/OE.25.010322} {\bibfield  {journal} {\bibinfo  {journal} {Optics
  Express}\ }\textbf {\bibinfo {volume} {25}},\ \bibinfo {pages} {10322}
  (\bibinfo {year} {2017}{\natexlab{b}})}\BibitemShut {NoStop}%
\bibitem [{\citenamefont {Mead}(1990)}]{Mead1990}%
  \BibitemOpen
  \bibfield  {author} {\bibinfo {author} {\bibfnamefont {C.}~\bibnamefont
  {Mead}},\ }\href {\doibase 10.1109/5.58356} {\bibfield  {journal} {\bibinfo
  {journal} {Proceedings of the IEEE}\ }\textbf {\bibinfo {volume} {78}},\
  \bibinfo {pages} {1629} (\bibinfo {year} {1990})}\BibitemShut {NoStop}%
\bibitem [{\citenamefont {Abediasl}\ and\ \citenamefont
  {Hashemi}(2015)}]{Abediasl2015}%
  \BibitemOpen
  \bibfield  {author} {\bibinfo {author} {\bibfnamefont {H.}~\bibnamefont
  {Abediasl}}\ and\ \bibinfo {author} {\bibfnamefont {H.}~\bibnamefont
  {Hashemi}},\ }\href {\doibase 10.1364/OE.23.006509} {\bibfield  {journal}
  {\bibinfo  {journal} {Optics Express}\ }\textbf {\bibinfo {volume} {23}},\
  \bibinfo {pages} {6509} (\bibinfo {year} {2015})}\BibitemShut {NoStop}%
\bibitem [{\citenamefont {Braitenberg}\ and\ \citenamefont
  {Schuz}(1998)}]{brsc1998}%
  \BibitemOpen
  \bibfield  {author} {\bibinfo {author} {\bibfnamefont {V.}~\bibnamefont
  {Braitenberg}}\ and\ \bibinfo {author} {\bibfnamefont {A.}~\bibnamefont
  {Schuz}},\ }\href@noop {} {\emph {\bibinfo {title} {Cortex: statistics and
  geometry of neuronal connectivity}}}\ (\bibinfo  {publisher} {Springer},\
  \bibinfo {address} {Berlin, Germany},\ \bibinfo {year} {1998})\BibitemShut
  {NoStop}%
\end{thebibliography}%

\end{document}